%% file: mikon_mk_paper_arxiv.tex
\documentclass[conference, a4paper]{IEEEtran}
\IEEEoverridecommandlockouts
\usepackage{cite}
\usepackage{amsmath,amssymb,amsfonts}
\usepackage{algorithmic}
\usepackage{graphicx}
\usepackage{textcomp}
\usepackage{xcolor}
\usepackage{hyperref}
\usepackage[
  separate-uncertainty = true,
  multi-part-units = repeat
]{siunitx}

\def\BibTeX{{\rm B\kern-.05em{\sc i\kern-.025em b}\kern-.08em
    T\kern-.1667em\lower.7ex\hbox{E}\kern-.125emX}}
\begin{document}
	
	\begin{titlepage}
		\begin{center}
			
			\Huge
			\textbf{Comparison of Extended and Unscented Kalman Filters Performance in a Hybrid BLE-UWB Localization System				
		}
			
			\vspace{0.5cm}
			\LARGE
			Accepted version
			
			\vspace{1.5cm}
			
			\text{Marcin Kolakowski}
			
			\vspace{.5cm}
			\Large
			Institute of Radioelectronics and Multimedia Technology
			
			Warsaw University of Technology
			
			Warsaw, Poland,
			
			contact: marcin.kolakowski@pw.edu.pl

			\vspace{2cm}

		\end{center}
		
		\Large
		\noindent
		\textbf{Originally presented at:}
		
		\noindent
		2020 23rd International Microwave and Radar Conference (MIKON), Warsaw, Poland, 2020
		
		\vspace{.5cm}
		\noindent
		\textbf{Please cite this manuscript as:}
		
		\noindent
M. Kolakowski, "Comparison of Extended and Unscented Kalman Filters Performance in a Hybrid BLE-UWB Localization System," 2020 23rd International Microwave and Radar Conference (MIKON), Warsaw, Poland, 2020, pp. 122-126, doi: 10.23919/MIKON48703.2020.9253854
		
		\vspace{.5cm}
		\noindent
		\textbf{Full version available at:}
		
		\noindent
		\url{https://doi.org/10.23919/MIKON48703.2020.9253854}

		%
		%
		
		\vfill
		
		\large
		\noindent
		© 2020 IEEE. Personal use of this material is permitted. Permission from IEEE must be obtained for all other uses, in any current or future media, including reprinting/republishing this material for advertising or promotional purposes, creating new collective works, for resale or redistribution to servers or lists, or reuse of any copyrighted component of this work in other works.
	\end{titlepage}

\title{Comparison of Extended and Unscented Kalman Filters Performance in a Hybrid BLE-UWB Localization System
\thanks{This work was financed by the National Centre for Research and Development, Poland under Grant AAL2/2/INCARE/2018.}
}

\author{
\IEEEauthorblockN{Marcin Kolakowski}
\IEEEauthorblockA{
\textit{Warsaw University of Technology}\\
\textit{Institute of Radioelectronics and Mult. Tech.} \\
Warsaw, Poland \\
m.kolakowski@ire.pw.edu.pl}
}

\maketitle

\begin{abstract}
The paper presents a comparison of performance of two Kalman Filters: extended Kalman filter (EKF) and unscented Kalman filter (UKF) in a hybrid Bluetooth-Low-Energy--ultra-wideband (BLE-UWB) based localization system. In the system, the user is localized primarily based on Received Signal Strength (RSS) measurements of BLE signals. The UWB part of the system is periodically used to improve localization accuracy by supplying the algorithm with measured UWB packets time difference of arrival (TDOA). The proposed scheme was experimentally validated using two algorithms: the EKF and the UKF. The localization accuracy of both algorithms is compared.
\end{abstract}


\input{tex/introduction}

\input{tex/scheme}

\input{tex/algorithms}

\input{tex/experiments}

\input{tex/conclusions}


\bibliographystyle{IEEEtran}
\bibliography{biblio2}

\end{document}

%% file: tex/introduction.tex
\section{Introduction}

One of the most popular technologies used for indoor positioning is Bluetooth Low Energy (BLE) \cite{cantonpaternaBluetoothLowEnergy2017a}. The main advantage of employing BLE standard in localization systems are its low energy requirements, which is an especially significant feature in case of worn devices (tags) as it allows them  to work for a much longer time without recharging than in case of systems using Wi-Fi or ultra-wideband technologies. It significantly reduces the cost of system maintenance in terms of time and money. Another benefit of BLE is its popularity. Almost everybody has a smartphone equipped with Bluetooth radio module, which opens a lot of opportunities to create easily accessible systems and services e.g. navigation inside museums or shopping malls. Additionally, thanks to the fact that Bluetooth is one of the most widely-spread radio technologies on the market, the cost of the system devices can be significantly reduced.

The typical BLE-based positioning systems allow for localization with accuracy in order of a few meters. The localization errors range from 1--2 to even 4--6 m depending on the number of anchors comprising the system infrastructure and the propagation conditions where the system is installed. Such accuracy might be insufficient for some cases, which need higher localization accuracy e.g. patient monitoring systems, which might require precise movement trajectory reconstruction. The localization accuracy of BLE-based systems can be achieved using various methods. One of them is hybrid localization.

In hybrid localization schemes, the location is calculated based on different types of measurement data obtained using one or many technologies. In the literature there are many examples of combining BLE with other technologies and techniques to improve system's localization accuracy and reliability. A very popular hybrid localization scheme combines radio based systems with the use of inertial measurement units \cite{jadidiRadioInertialLocalizationTracking2018a}, \cite{huangHybridMethodImprove2019a}. In \cite{jadidiRadioInertialLocalizationTracking2018a} the inertial sensors and a magnetometer are used for dead-reckoning-based localization, in places, which are not covered with BLE part of the system. The solution presented in \cite{huangHybridMethodImprove2019a} uses the additional sensors to estimate person's movement direction and gait length, which allows to improve localization accuracy.

Another examples of hybrid localization schemes combine results obtained using different radio technologies. In \cite{antevskiHybridBLEWiFi2016b} a fingerprinting-based system using BLE and Wi-Fi is presented. In the system, two different radio maps for both technologies are used, which allows to detect local power level fluctuations and avoid large localization errors. BLE-based localization systems can also be supported with UWB. In \cite{kolakowskiKalmanFilterBased2017a} the UWB system is periodically used to improve localization accuracy of BLE-based system.

The measurement results coming from the system devices might be processed by a wide range of algorithms. The popular choice are various versions of the Kalman filter. Since the dependence of typically conducted measurement results (received signal strength, propagation time) on the localized object's location is nonlinear, most of the systems utilize either the extended Kalman filter (EKF) or the unscented Kalman filter (UKF). The EKF handles the nonlinearity by employing simple linearization, whether the UKF uses a more accurate unscented transform.  Both of these algorithms have been tested and compared in various localization \cite{giannitrapaniComparisonEKFUKF2011} and data fusion applications \cite{xiaComparisonCentralisedScaled2016}. The results of the above studies have shown that, in most cases, the UKF allows to achieve better results than the EKF. The object of the presented study is to compare the accuracy of those filters in a hybrid localization scheme combining BLE and UWB technologies. 

The presented research is a direct continuation of works published in \cite{kolakowskiKalmanFilterBased2017a}. In the earlier scheme, the user was primarily localized using Bluetooth Low Energy based part of the system. The UWB-based part was used incidentally. The data from both parts were processed using an EKF-based algorithm. The results of the first experiments performed using and UWB-based system and BLE Texas Instruments evaluation boards have shown that even relatively rare (once every 4 seconds) use of UWB technology, allows to reduce localization errors.

In the study presented in the paper, the above hybrid localization concept was tested in a fully hybrid BLE-UWB localization system, which was developed at Warsaw University of Technology. The main goal of the experiments was to determine, whether employing a bit more complicated UKF in the system allows to achieve better localization results in respect to the basic EKF. The structure of the paper is as follows. Section \ref{sec:scheme} includes the description of the proposed localization scheme. The algorithms used in the research are described in Section \ref{sec:algorithms}. The results of the performed experiments are presented in Section \ref{sec:experiments}.

%% file: tex/scheme.tex
\section{Localization scheme}
\label{sec:scheme}

\begin{figure}[b]
\centering
\centerline{\includegraphics[width=.8\linewidth]{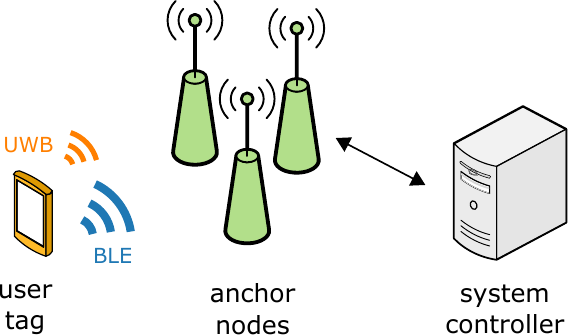}}
\caption{The proposed hybrid localization scheme}
\label{fig:scheme}
\end{figure}

The proposed hybrid localization scheme is illustrated with a diagram presented in Fig. \ref{fig:scheme}. 
In the proposed localization scheme, the system consists of three parts:
\begin{itemize}
\item a localized tag equipped with BLE and UWB radio modules,
\item a system infrastructure comprising a set of synchronized anchors capable of receiving BLE and UWB signals,
\item a system controller.
\end{itemize}

In the proposed system, during its routine work, the tag transmits both BLE and UWB packets. The modules are used with different intensity. The tag transmits 3 BLE packets per second, whereas UWB module is used at a lower, programmable rate. Such approach allows to reduce energy consumption of the tag device in respect to the situation, in which the modules worked at the same rate (energy requirements of BLE system part are much lower compared to its UWB counterpart).

The transmitted packets are received by the anchors, which measure BLE signal level and UWB packets times of arrival. The measurement results are then relayed to the system controller, which processes them to calculate user's location using a hybrid RSS-TDOA--based algorithm.

%% file: tex/algorithms.tex
\section{Localization algorithms}
\label{sec:algorithms}
\subsection{System model}
\label{sec:model}
In the proposed system, a localized user (or a tag) is modeled as a dynamic system, which state at a given moment $k$ is described with a state vector
\begin{equation}
x_k = \begin{bmatrix}
x&v_x&y&v_y\\
\end{bmatrix}\label{eq:state_vec}
\end{equation} 
including user's coordinates $x$, $y$ and velocity expressed as its components in perpendicular directions $v_x$, $v_y$. The uncertainty associated with the following features is expressed with the state vector covariance matrix $P_k$ The changes occurring in the system (movement of the localized person) are described using the following discrete-time model:
\begin{equation}
x_{k}=Fx_{k-1} + w_{k-1}\label{eq:TUmodel}
\end{equation}
\begin{equation}
F = \begin{bmatrix}
F_t&0_{2x2}\\
0_{2x2}&F_t\\
\end{bmatrix}, \>  F_t = \begin{bmatrix}
1&\Delta t\\
0&1\\
\end{bmatrix}
\end{equation}
\begin{equation}
Q = \begin{bmatrix}
Q_t&0_{2x2}\\
0_{2x2}&Q_t\\
\end{bmatrix},
Q_{t} = 
\begin{bmatrix}
\frac{1}{4}\Delta t^4&\frac{1}{2}\Delta t^3\\\frac{1}{2}\Delta t^3 & \Delta t^2\end{bmatrix} \sigma_{ax}^2
\end{equation}
where $F$ is the state transition matrix containing movements of motion chosen with accordance to the Discrete White Noise acceleration model DWNA \cite{bar-shalomEstimationApplicationsTracking2001a}, which is a uniform linear motion model treating the acceleration as discrete white noise $w_{k-1}$ of covariance $Q$. The variance of the process noise depends on acceleration variance $\sigma_{ax}^2$ and system update period $\Delta t$.

In the system, the anchors measure BLE signal levels and UWB packets time of arrival. The sensor model of the system is therefore:
\begin{multline}
z = h(x_k) +v_k = [P_1(x_k) \hdots P_m(x_k)\\T_1(x_k) \hdots T_n(x_k)] +v_k\label{eq:sensormodel}
\end{multline}
Where $P_i(x)$ and $T_i(x)$ are the measured BLE signal level and  the TDOA value calculated for UWB packets time of arrival respectively, $v_k$ is the measurement noise of covariance $R_k$ The received signal levels are modeled with a log-distance path-loss model:
\begin{equation}
P_i(x) = P_{i0} -10\gamma\log\frac{d_{i}(x)}{d_0}\label{eq:rss}
\end{equation}
where $P_{i0}$  is the power received by the anchor $i$ when the tag is at reference distance $d_0$ (1 m), $d_i(x)$ is the distance between the anchor and the tag and $\gamma$ is the path-loss exponent. The TDOAs are calculated as a difference between the propagation times between two anchors $k$,$l$ and a tag:
\begin{equation}
T_{kl}(x) = \frac{1}{c} (|x - s_k|-|x - s_l|)\label{eq:tdoa}
\end{equation}
where $c$ is the speed of light and $x$, $s_k$ and $s_l$ are the locations of the tag and both anchors respectively.

\subsection{Localization algorithms}

\subsubsection{Extended Kalman Filter}
The extended Kalman filter (EKF) \cite{mohindersgrewalKalmanFilteringTheory2015a} is a well established algorithm used for state estimation in dynamic systems. It is described with a following set of equations:
\begin{equation}
\hat{x}_{k(-)}=F\hat{x}_{k-1(+)}\label{eq:TU}
\end{equation}
\begin{equation}
P_{k(-)}=FP_{k-1(+)}F^T+Q\label{eq:CTU}
\end{equation}
\begin{equation}
\hat{x}_{k(+)}=\hat{x}_{k(-)}+K_k\left(z_k-h_k(\hat{x}_{k(-)})\right)\label{eq:MU}
\end{equation}
\begin{equation}
P_{k(+)}=\left(I-K_kH_k^T\right)P_{k(-)}\label{eq:CMU}
\end{equation}
\begin{equation}
H_k=\frac{\partial h_k(x)}{\partial x}\Bigr|_{x=\hat{x}_{k(-)}}\label{eq:ekf_lin}
\end{equation}
\begin{equation}
K_k = P_{k(-)}H_k^T\left(H_kP_{k(-)}H_k^T+R_k\right)^{-1}\label{eq:gain}
\end{equation}
where (\ref{eq:TU}-\ref{eq:CTU}) comprise the time-update phase, in which the current state vector value is estimated based on the value in the previous moment $k-1$ and the movement model described in Section \ref{sec:model}. Equations (\ref{eq:MU}-\ref{eq:gain}) describe the measurement update phase, in which the previously obtained state vector estimate is updated based on the measurement results. The extent to which the EKF result will be affected by measurement results depends on the Kalman gain $K_k$. As the sensor model (\ref{eq:sensormodel}-\ref{eq:tdoa}) is non-linear, calculating $K_k$ is preceded by its linearization in the predicted point (\ref{eq:ekf_lin}).

\subsubsection{Unscented Kalman Filter}
The linearization of the sensor model implemented in the EKF (\ref{eq:ekf_lin}) might not be accurate enough in case of highly non-linear functions. The unscented Kalman filter (UKF) \cite{wanUnscentedKalmanFilter2000} avoids computing Jacobian matrix by using the unscented transformation, which is a method allowing to calculate the statistics (mean, covariance) of a random variable being a result of a non-linear transformation.

As in the proposed system model, the state transition (movement model) is linear, the time update phase of the UKF is the same as in the EKF (\ref{eq:TU}-\ref{eq:CTU}). The unscented transformation is only applied in the measurement-update phase.

The unscented transformation is performed in two steps. In the first step a set of sigma points ($\mathcal{X}$) and corresponding weights is created in a following way:
\begin{equation}
\mathcal{X}_0 = \hat{x}_{k(-)}
\end{equation}
\begin{equation}
\mathcal{X}_i = \hat{x}_{k(-)} + (\sqrt{(L+\lambda)P_{k(-)}})_i \qquad i=1,\hdots,L
\end{equation}
\begin{equation}
\mathcal{X}_i = \hat{x}_{k(-)} - (\sqrt{(L+\lambda)P_{k(-)}})_i \qquad i=L+1,\hdots,2L
\end{equation}
\begin{equation}
\lambda = \alpha^2(L+\kappa)-L
\end{equation}
\begin{equation}
W_0^{(m)} = \lambda/(L+\lambda)
\end{equation}
\begin{equation}
W_0^{(c)} = \lambda/(L+\lambda) + (1-\alpha^2+\beta)
\end{equation}
\begin{equation}
W_i^{(m)} = W_i^{(c)} =  \frac{1}{2(L+\lambda)}
\end{equation}
The first sigma point $\mathcal{X}_0$ is the mean, the next ones $\mathcal{X}_i$ are the columns of the root of scaled state vector covariance matrix. There are two sets of weights  $W^{(m)}$ and $W^{(c)}$, which will be used for mean and covariance reconstruction respectively. The effect of the unscented transformation depends on a set of parameters: $L=4$ which is a dimension of the state vector, $\alpha$ which is a scaling factor determining the spread of sigma points around the mean, $\kappa$ typically set to 0 and $\beta$ which in case of Gaussian distribution equals 2. The second step is the estimation of the measurement results by transforming the obtained sigma points using the sensor model and reconstructing  the mean value and the covariance matrix:
\begin{equation}
\mathcal{X}_i = h_k(\mathcal{X}_i) \qquad i=0,\hdots,2L
\end{equation}
\begin{equation}
\hat{z}_k = \sum_{0}^{2L} W_i^{(m)}\mathcal{Z}_i
\end{equation}
\begin{equation}
P_{z,z} = \sum_{0}^{2L} W_i^{(c)}[\mathcal{Z}_i - \hat{z}_k ][\mathcal{Z}_i - \hat{z}_k ]^T + R_k
\end{equation}

The rest of the measurement-update phase is performed in the same way as in case of the EKF, but the individual values are calculated in a slightly different way.

\begin{equation}
P_{x,z} = \sum_{0}^{2L} W_i^{(c)}[\mathcal{X}_i - \hat{x}_k(-) ][\mathcal{Z}_i - \hat{z}_k ]^T
\end{equation}
\begin{equation}
K = P_{x,z}P_{z,z}^{-1}
\end{equation}
\begin{equation}
\hat{x}_{k(+)}=\hat{x}_{k(-)}+K_k\left(z_k- \hat{z}_k\right)\label{eq:MUKF}
\end{equation}
\begin{equation}
P_{k(+)}=P_{k(-)}-KP_{z,z}K^T\label{eq:CMUkf}
\end{equation}

%% file: tex/experiments.tex
\section{Experiments}
\label{sec:experiments}

\begin{figure}[hb]
\centering
\centerline{\includegraphics[width=.7\linewidth]{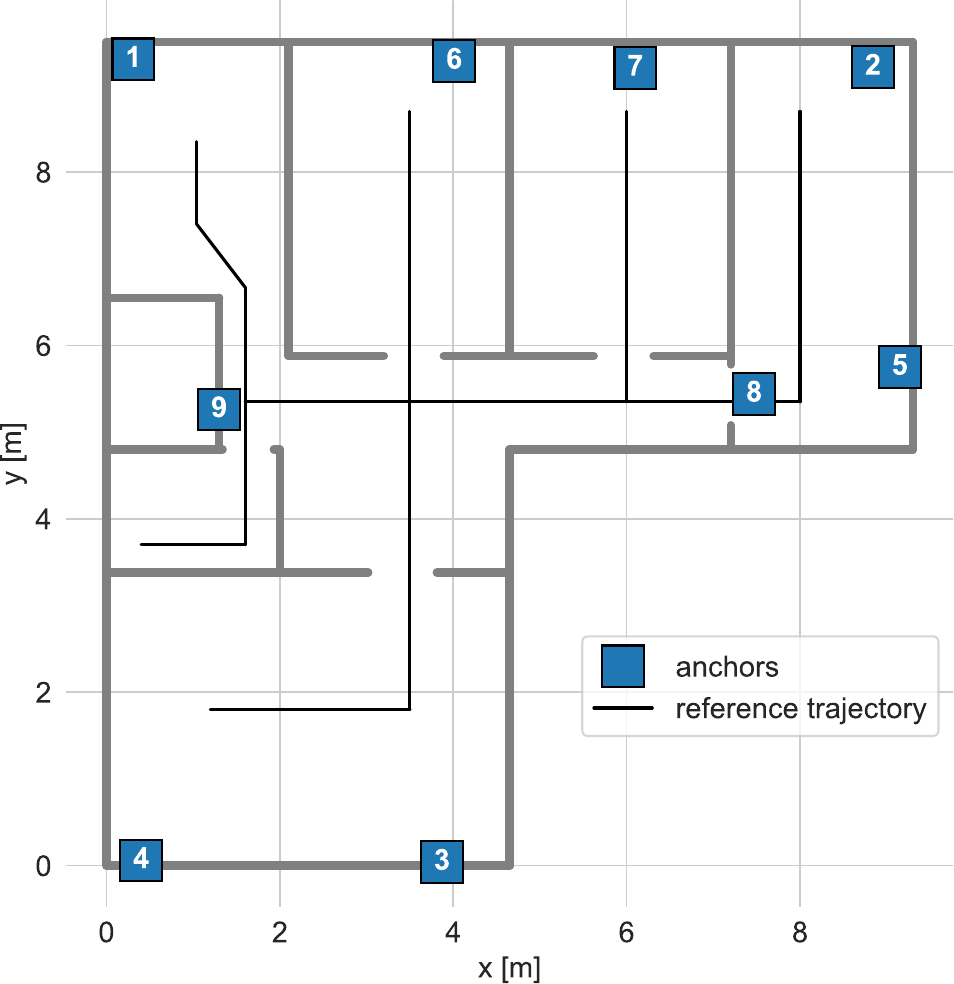}}
\caption{The experiment site and system infrastructure layout}
\label{fig:layout}
\end{figure}

The proposed hybrid localization scheme was evaluated using the algorithms proposed in Section \ref{sec:experiments}. The experiments were performed in a fully furnished apartment consisting of four rooms and a kitchen. The plan of the apartment and the layout of the system infrastructure employed in the experiment are presented in Fig. \ref{fig:layout}. 

\begin{figure}[hb]
\centering
\centerline{\includegraphics[width=.8\linewidth]{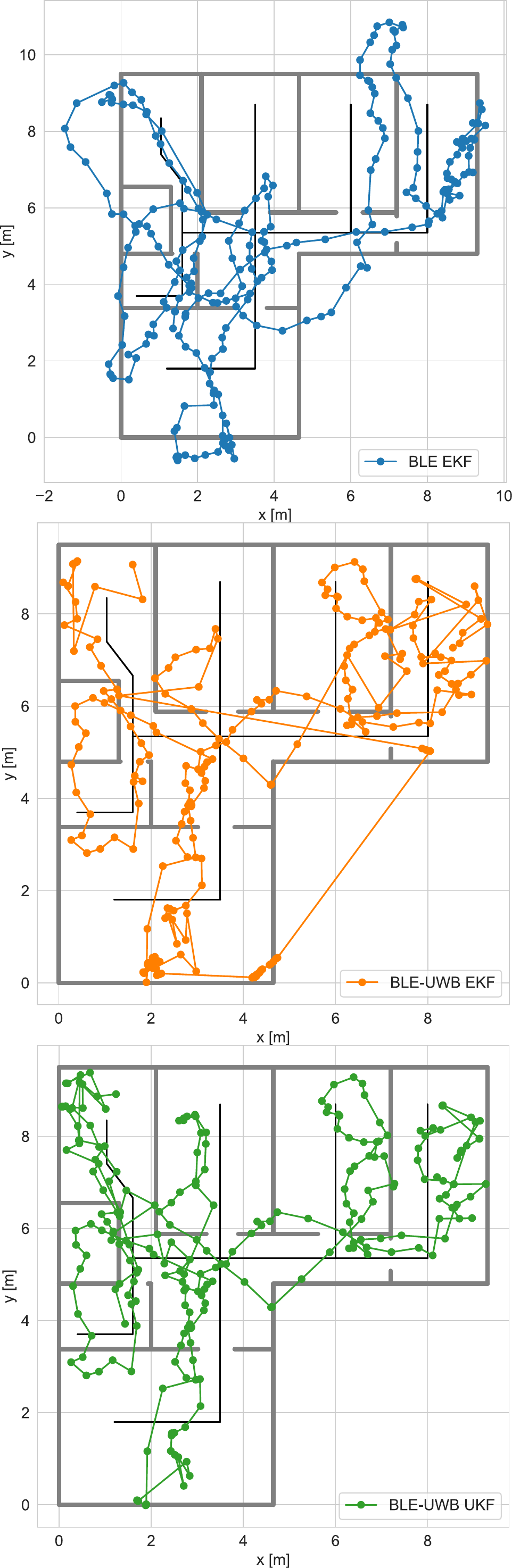}}
\caption{Exemplary localization results (3 UWB packets per second)}
\label{fig:results}
\end{figure}
The system used in the experiment was developed at WUT during one of our earlier projects \cite{kolakowskiLocalizationSystemSupporting2020}. The configuration used in the tests consisted of 9 hybrid BLE-UWB anchors, which were placed in the apartment. The anchors were evenly distributed, as in each of the rooms at least one device was placed. During the tests, four different persons were asked to wear a tag and walk along a predefined reference path covering all of the rooms in the apartment. The gathered results were then processed with different Kalman-filter-based algorithms to assess the system's accuracy and verify, which version of the KF better suits the proposed scheme.

In the experiment the UWB part of the system worked at the same rate as the BLE one (the tag was transmitting 3 BLE and 3 UWB packets per second). The gathered UWB results were then decimated in the post-processing to assess the influence on UWB use rate on localization accuracy. The exemplary localization results are presented in Fig. \ref{fig:results}

The results obtained using solely BLE are least accurate. In case of EKF and UKF the results are very similar although there are some significant differences. For example, using UKF it was possible to observe walking into the second room from the left, whereas it was not visible in BLE-based and EKF-based localization results.

\begin{figure}[t]
\centering
\centerline{\includegraphics[width=.8\linewidth]{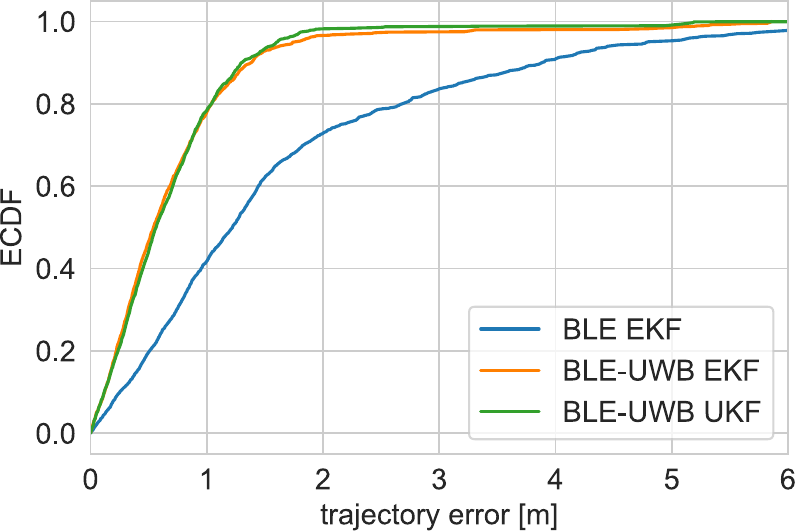}}
\caption{Trajectory error ECDF (3 UWB packets per second)}
\label{fig:ecdf1}
\end{figure}

\begin{figure}[t]
\centering
\centerline{\includegraphics[width=.8\linewidth]{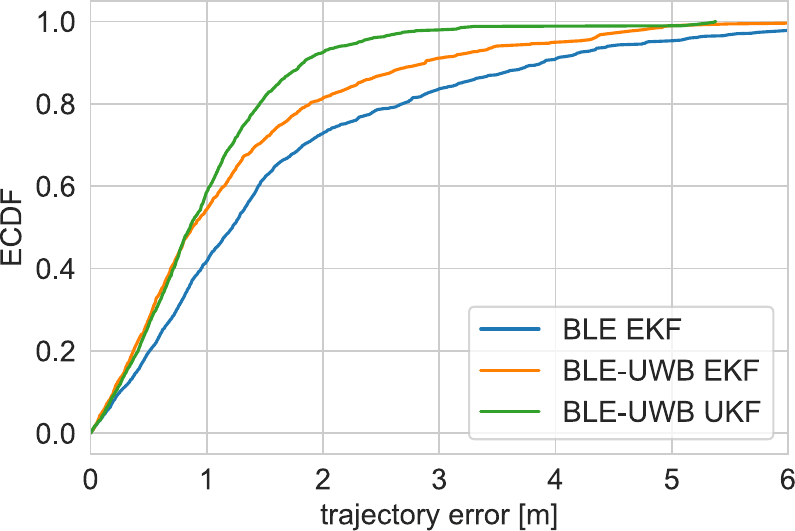}}
\caption{Trajectory error ECDF (one UWB packet every two seconds)}
\label{fig:ecdf2}
\end{figure}

An easier comparison can be done based on the empirical cumulative distribution functions plotted for trajectory error. In the study, the trajectory error is defined as the shortest distance of the localized point to the trajectory line. The ECDF plotted for the results obtained when the tag transmits three UWB packets per second and one UWB packet every two seconds are presented in Fig.\ref{fig:ecdf1} and Fig.\ref{fig:ecdf2} respectively. 

In case of high UWB rate, there is no significant difference between trajectory errors of locations obtained using EKF and UKF. However, when the UWB packet transmission rate is reduced, the difference in favor of the UKF is clearly visible. Only 10\% of the UKF have a trajectory error higher than 2 m. In case of the EKF, such errors are present in 20\% of the results.

%% file: tex/conclusions.tex
\section{Conclusions}
In the paper, the extended and unscented Kalman filters were compared in terms of their performance in a hybrid BLE-UWB localization system. In the system the users are primarily localized based on the RSS measured for BLE signals. The UWB part of the system is used incidentally. Such approach allows to improve localization accuracy, while keeping energy usage at moderate levels.

The proposed system concept was experimentally tested. During the experiments four walking persons were localized using EKF and UKF. The comparison of results have shown that, when the number of the UWB packets transmitted by the tag is high, there is no significant difference in EKF and UKF performance. However, when the working rate of the UWB part is reduced, the results obtained with the UKF are superior.